\documentclass[prd,twocolumn,twoside,nofootinbib,preprintnumbers,superscriptaddress,showpacs,showkeys]{revtex4-1}

\usepackage{amsmath}
\usepackage{amssymb}  
\usepackage{mathrsfs} 
\usepackage{graphicx}
\usepackage{dcolumn}
\usepackage{color}
\usepackage[breaklinks=true]{hyperref}

\usepackage{wick}

\def\esl{E \hspace{-0.6em}/\hspace{0.2em}}

\def\gsim{\raise0.3ex\hbox{$\;>$\kern-0.75em\raise-1.1ex\hbox{$\sim\;$}}}
\def\lsim{\raise0.3ex\hbox{$\;<$\kern-0.75em\raise-1.1ex\hbox{$\sim\;$}}}

\begin{document}

\preprint{IFIC/19-08, IFT-UAM/CSIC-19-23, FTUAM-19-7}

\title{Proton decay at 1-loop}

\author{Juan Carlos Helo}
\email{jchelo@userena.cl}
\affiliation{Departamento de F\'{i}sica,
Facultad de Ciencias, Universidad de La Serena,
Avenida Cisternas 1200, La Serena, Chile}

\author{Martin Hirsch}
\email{mahirsch@ific.uv.es}
\affiliation{AHEP Group, Instituto de F\'{i}sica Corpuscular
- CSIC/Universitat de Val\`{e}ncia Edificio de Institutos
de Paterna, Apartado 22085, E–46071 Val\`{e}ncia, Spain}

\author{Toshihiko Ota}
\email{toshihiko.ota@uam.es}
\affiliation{Departamento de F\'{i}sica Te\'{o}rica and
Instituto de F\'{i}sica Te\'{o}rica, IFT-UAM/CSIC,
Universidad Aut\'{o}noma de Madrid,
Cantoblanco, 28049 Madrid, Spain}

\begin{abstract}

Proton decay is usually discussed in the context of grand unified theories.
However, as is well-known, in the standard model effective theory  
proton decay appears in the form of higher dimensional non-renormalizable
operators.
Here, we study systematically the 1-loop decomposition of the
$d=6$ $B+L$ violating operators. We exhaustively list the possible
1-loop ultra-violet completions of these operators and discuss that,
in general, two distinct classes of models appear.
Models in the first class need an additional symmetry in order to
avoid tree-level proton decay. These models necessarily contain
a neutral particle, which could act as a dark matter candidate.
For models in the second class the loop contribution dominates
automatically over the tree-level proton decay, without the need
for additional symmetries.
We also discuss possible phenomenology of two example models,
one from each class, and their possible connections to neutrino masses,
LHC searches and dark matter.
\end{abstract}



\maketitle

\section{Introduction}
\label{sect:intro}

While searches for proton decay so far have yielded only lower bounds
on the lifetime of various possible decay modes~\cite{Sussman:2018ylo,TheSuper-Kamiokande:2017tit,Miura:2016krn,Takhistov:2015fao,TheKamLAND-Zen:2015eva,Gustafson:2015qyo,Takhistov:2014pfw,Abe:2014mwa,Abe:2013lua,Tanabashi:2018oca},
future large volume detectors, such as Hyper-Kamiokande \cite{Abe:2018uyc},
DUNE \cite{Abi:2018dnh}
and JUNO \cite{An:2015jdp},
or more speculative multi-megaton proposals such as
TITAND \cite{Suzuki:2001rb,Kistler:2008us},
MEMPHYS \cite{deBellefon:2006vq} or
MICA \cite{Boser:2013oaa,Cowen:2013zz}
offer a good chance to finally
discover this ultra-rare process.
Although nucleon decay processes are usually discussed
in the context of grand unified theories (GUTs), see e.g.,
Refs.~\cite{Georgi:1974sy,Langacker:1980js,Sakai:1981pk,Weinberg:1981wj,Dimopoulos:1981dw,Ellis:1981tv,Nath:1985ub,Hisano:1992jj,Lucas:1996bc,Goto:1998qg,Murayama:2001ur,Hisano:2013exa,Ellis:2015rya},
they can arise in many models.
For a review on baryon number violation,
see for example Ref.~\cite{Babu:2013jba}.
Motivated by the expected improvements in nucleon decay searches,
here we study proton decay generated at the 1-loop level.

In the standard model baryon and lepton number violation arises at the
non-renormalizable level. At the level of mass dimension five ($d=5$),
there is only one operator, the famous Weinberg
operator~\cite{Weinberg:1979sa}, corresponding to
Majorana neutrino masses ($\Delta L=2$, $\Delta B=0$).
At $d=6$ there are already five independent operators, which have $\Delta
B=\Delta L=1$ (but $\Delta (B-L)=0$)
\cite{Weinberg:1979sa,Wilczek:1979hc,Abbott:1980zj}.
All $d=6$ operators lead to two-body proton decays, such as
$p \to \pi^0 + e^+$, $p \to \pi^+ + \bar{\nu}$ or $p \to K^+ + \bar{\nu}$. 

GUT models predict proton decay to occur at tree-level
\cite{Georgi:1974sy,Langacker:1980js}.
For coefficients of order ${\cal O}(1)$, the current experimental bounds
then imply a lower limit on the scale of baryon number violation
(for $d=6$ operators) of order
$\Lambda \sim {\rm (few)}\hskip1mm{\cal O}(10^{15})$ GeV,
which is far out of reach of any foreseeable accelerator experiment.
This simple picture changes drastically, if proton decay is induced
by higher dimensional operators and/or at loop level.
The decay rate for a $k$-body $n$-loop proton/neutron decay induced
by a $d$-dimensional operator can be very roughly estimated to be: 
\begin{equation}\label{eq:pdec}
\frac{1}{\tau} \sim \frac{{\cal C}^{2}}{f[k]} \Big(\frac{1}{16 \pi^2}\Big)^{2 n}
\Big(\frac{m_p}{\Lambda}\Big)^{2(d-6)}\frac{m_p^5}{\Lambda^4}
\end{equation}
Here,
$f[k] \equiv4\left(4\pi\right)^{2k-3}\left(k-1\right)!
\left(k-2\right)!$ estimates the phase space volume available to the
decay products for massless final state particles~\cite{Fonseca:2018ehk}.
The constant ${\cal C}$ is the coefficient of the effective interaction
that induces the proton decay process, which contains products
of couplings that appear in the ultra-violet models given
at the scale $\Lambda$.
Note that ${\cal C}$ can be small compared to one, depending on the
model, see below.
Obviously, to obtain decay 
rates within future experimental sensitivities much lower scales
$\Lambda$ are needed for $k \gg 2$, $n \gg 0$ and/or $d \gg 6$.

Probably for this reason, not many studies on
higher-dimensional proton decay operators can be found in the
literature.
For $d=7$ operators see, for example,
Refs.~\cite{Babu:2012iv,Lehman:2014jma,Bhattacharya:2015vja}.
For operators with $d=9$ and higher see
Refs.~\cite{ODonnell:1993kdg,Hambye:2017qix,Fonseca:2018ehk}.
In particular, Ref.~\cite{Fonseca:2018ehk} discusses $\Delta L=3$
proton decay from operators up to $d=13$, where current experimental
sensitivities correspond to new physics scales $\Lambda \lsim $TeV,
even for couplings as large as order ${\cal O}(1)$.\footnote{%
The complete list of high-$d$ operators can easily be obtained with
{\tt Sym2Int}~\cite{Fonseca:2017lem}.}
The authors of Ref.~\cite{deGouvea:2014lva}
listed the higher-mass-dimensional $B-L$-violating effective operators
in a GUT model and discussed the relations between neutrino masses
and the nucleon decays induced by the effective operators.

Even less work has been done so far for loop-induced proton decay.
Perhaps the best-known example for it
is supersymmetric (SUSY) GUTs, see for example
the review~\cite{Babu:2013jba}.
Here, the importance of the loop stems from the fact that
the decay amplitude is proportional to 
$(\Lambda_{\rm GUT}\Lambda_{\rm SUSY})^{-1}$ instead of
$\Lambda_{\rm GUT}^{-2}$ 
as (for tree-level contributions) in non-SUSY GUT models.

In this paper, we exhaustively list the possible high-energy
completions of the proton decay operators with $d=6$ at
the 1-loop level.
We also calculate group theoretical factors and define
the 1-loop integrals, which appear in the reproduction
of the proton decay operators from their decompositions.
From these lists one can immediately estimate the rate
of proton decay, once a (proto-)model is specified.
For masses of the mediators at the TeV scale, we find that
the couplings $Y$ entering the proton decay rate should be
of order $Y < \mathcal{O}(10^{-6})$.\footnote{%
The coefficient ${\cal C}$ in Eq.~\eqref{eq:pdec} is
${\cal C} \propto Y^4$ in 1-loop $d=6$ models.}
This opens up the possibility that the
charged/coloured mediator fields live long on the time scale of
collider experiments, yielding particular signals at the LHC.

We divide the different models, found in our lists, into two
sub-classes.
Models in the first class require an additional symmetry to avoid
tree-level proton decay.
It is straightforward to introduce some extra symmetry in these cases,
for example a $Z_2$, that guarantees that proton decay appears only at
the 1-loop (and higher) level.
In this class of models the lightest loop particle is then
necessarily stable and thus can serve as a candidate for the dark
matter.
In the second class one finds models,
in which the loop-induced $d=6$ decay is automatically the leading
contribution to proton decay, despite the existence of tree-level
decay modes.
The reason for this counter-intuitive behaviour is simply
that for models in the second class, tree-level proton decay 
appear only at the level of higher-dimensional effective operators.

We then discuss two example models, one from each model class, in more
details.
In Model-I, neutrino masses, dark matter and proton decay
are all related.
Majorana neutrino masses are generated using the scotogenic
loop~\cite{Ma:2006km} and the same $Z_2$ that stabilizes
the dark matter guarantees that proton decay occurs only at the 1-loop
level.
The coloured mediators of proton decay, if at the TeV scale,
can be produced at the LHC and will decay to jets, leptons
and the dark matter candidate.
These missing energy signals, possibly associated with charged tracks
from heavy ionizing particles, are reminiscent of those discussed
in the context of SUSY.
Thus, one can use different existing searches at the LHC to derive
constraints on the model.
Also, since the model generates neutrino masses at 1-loop, one can
constrain its parameters using searches for lepton flavour violation,
such as $\mu\to e\gamma$ and others.

In our Model-II we do not impose any beyond the SM symmetry.
Thus, there are no stable, heavy particles.
Signals for searches at the LHC are therefore different from
those discussed for Model-I.
In particular, there are final states with no missing energy involved.	
For this model, we also show how tree-level proton decay will 
appear and is suppressed in models in this class.
For the particular case of Model-II, the final state for proton decay
is caused by a tree $d=12$ operator and is 5-body. 
The expected partial half-lives for these 
modes are therefore orders of magnitude larger than those of the 
1-loop induced 2-body decays.

The rest of this paper is organized as follows. In Section II 
we will discuss the $d=6$ operators and their 1-loop decomposition.
Section III then presents and discusses our two example models, 
before we conclude in Section IV.
Some more technical aspects 
for the 1-loop decomposition are given in the appendix.

\section{Proton decay operators at 1-loop}

The effective operators which lead to proton decay were
already listed in
Refs.~\cite{Weinberg:1979sa,Wilczek:1979hc,Abbott:1980zj}\footnote{%
See refs.~\cite{Alonso:2014zka,Helo:2018bgb}
for the $d=6$ operators for proton decay with a SM singlet fermion
(sterile neutrino, aka right-handed neutrino).}:
\begin{align}
 \mathcal{O}_{1} =& [du][QL],
 \label{eq:O1}
 \\
 \mathcal{O}_{2} =& [QQ][ue],
 \label{eq:O2}
 \\
 \mathcal{O}_{3} =& [QQ]_{\bf 1}[QL]_{\bf 1},
 \label{eq:O3}
 \\
 \mathcal{O}_{4} =& [QQ]_{\bf 3}[QL]_{\bf 3},
 \label{eq:O4}
 \\
 \mathcal{O}_{5} =& [du][ue],
 \label{eq:O5}
\end{align}
where the subscripts ${\bf 1}$ and ${\bf 3}$
in Eqs.~\eqref{eq:O3} and \eqref{eq:O4}
indicate the electroweak $SU(2)$ representation of
the bilinears of the fermions.
The contraction of all the indices on the operators is
explicitly shown in Appendix.
%

\begin{figure}[t]
\unitlength=1cm
\begin{picture}(5.8,3)
\put(0,0){\includegraphics[width=5.5cm]{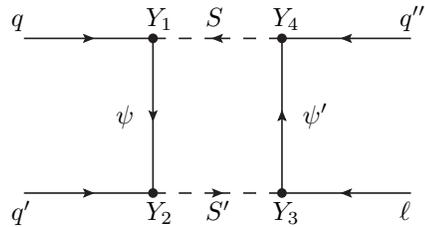}}
  \end{picture}
 \caption{Box diagram for decomposition of the $d=6$ proton
 decay operators. The direction of the arrows represents
 the flow of the particle number (not chirality).
 We put the lepton field $\ell$ always in the lower right corner $(Y_{3})$
 in all decompositions, i.e.,
 The $SU(3)$ structure is common in all decompositions.}
 \label{Fig:Box}
\end{figure}
We are particularly interested in the effective operators
which arise from 1-loop diagrams.
We decompose the effective operators
Eqs.~\eqref{eq:O1}-\eqref{eq:O5}
and list the necessary mediators
and the interactions.
There are two types of topologies for $d=6$ 1-loop diagrams:
triangle and box.
However, the decompositions based on the triangle diagram
allow to have the same effective operator at the tree level.\footnote{%
Forbidding a Yukawa interaction that mediates a $d=6$
proton decay operator at the tree level by a symmetry
and allowing the soft breaking of the symmetry by the mass term
of a mediator field, one can induce a triangle diagram
solely at the loop level.
For more discussions on the realizations,
see Ref.~\cite{Bonnet:2012kz}.
In this study, we do not pursue the possibility of such a setup
with a symmetry and its soft breaking.
}
Therefore, we concentrate on the decompositions with
the box diagram.
In the decomposition, 
we distribute the quarks and the lepton of an effective operator
to the outer legs of the box diagram shown in Fig.~\ref{Fig:Box}
in all possible ways,
and identify the SM gauge charges of the mediator fields,
$\psi$, $S$, $\psi'$, and $S'$.
In the current study, we restrict ourselves
to the decompositions with scalar and fermion mediators
and 
do not introduce a vector mediator which may require
an extension of the SM gauge symmetries
and their spontaneous breaking to the SM.
In short, we introduce the following Yukawa interactions,
\begin{align}
 \mathscr{L}
 =&
 Y_{1}
 \overline{q^{c}}
 \psi^{c}
 S
 +
 Y_{2}
 \overline{\psi^{c}} q'
 S'^{\dagger}
 +
 Y_{3}
 \overline{\psi'}
 \ell
 S'
 +
 Y_{4}
 \overline{q''^{c}}
 \psi'
 S^{\dagger}
 \nonumber
 \\
 &+
 {\rm H.c.},
 \label{eq:L-decom-general}
\end{align}
where $q$, $q'$, and $q''$ are the quark fields
($Q$, $u$, or $d$),
and
$\ell$ is the lepton field
($L$ or $e$) in an effective operator.
The mediator fields, 
$\psi$ and $\psi'$ for fermions
and $S$ and $S'$ for scalars, 
are assigned as shown in Fig.~\ref{Fig:Box}.
The mass terms for the mediator fields must also be included in the
Lagrangian, although they are not explicitly written
in Eq.~\eqref{eq:L-decom-general}.
Later we will discuss the phenomenology of mediator fields, assuming
that the masses $M_{\psi}$, $M_{S}$, $M_{\psi'}$, and $M_{S'}$ are at
the TeV scale.

The colour $SU(3)$ structure of the box diagram Fig.~\ref{Fig:Box}
is common in all decompositions,
and the possible ways to assign the $SU(3)$ charges
to the mediators are listed in Tab.~\ref{Tab:SU3-decom}.
Here we assume that a mediator takes one 
of ${\bf 1}$,
${\bf 3}$, $\overline{\bf 3}$,
${\bf 6}$, $\overline{\bf 6}$,
and ${\bf 8}$ representations under the $SU(3)$
transformation and do not pursue the possibility
of decompositions with a mediator whose
representation is higher than ${\bf 8}$.
\begin{table}[t]
\begin{tabular}{cccccc}
 \hline \hline
 & \multicolumn{4}{c}{Mediators}
     &
 \\
 & $\psi$ & $S$ & $\psi'$ & $S'$ &
 $SU(3)$ coeff.
 \\
 \hline
 \#1 &${\bf 1}$ & $\overline{\bf 3}$ & ${\bf 3}$ & ${\bf 3}$
 & $-1$
	     \\
 \hline
 \#2 &${\bf 3}$ & ${\bf 1}$ & $\overline{\bf 3}$ & $\overline{\bf 3}$
 & $1$
		 \\
 \#3 &${\bf 3}$ & ${\bf 8}$ & $\overline{\bf 3}$ & $\overline{\bf 3}$
 & $-\frac{8}{3}$
		 \\
 \#4 &${\bf 3}$ & ${\bf 8}$ & ${\bf 6}$ & ${\bf 6}$
 & $4$
		 \\
 \hline
 \#5 & $\overline{\bf 3}$ & ${\bf 3}$ & ${\bf 1}$ & ${\bf 1}$
 & $1$
		 \\
 \#6 & $\overline{\bf 3}$ & ${\bf 3}$ & ${\bf 8}$ & ${\bf 8}$
 & $-\frac{8}{3}$
		 \\
 \#7 & $\overline{\bf 3}$ & $\overline{\bf 6}$ & ${\bf 8}$ & ${\bf 8}$
 & $-4$
 \\
 \hline
 \#8 & ${\bf 6}$ & ${\bf 3}$ & {\bf 8} & ${\bf 8}$
 & $4$
 \\
 \hline
 \#9 & $\overline{\bf 6}$ & ${\bf 8}$ & $\overline{\bf 3}$ & $\overline{\bf
 3}$
 & $-4$
		 \\
 \hline
 \#10 & ${\bf 8}$ & $\overline{\bf 3}$ & ${\bf 3}$ & $ {\bf 3}$
 & $\frac{8}{3}$
		 \\
 \#11 & ${\bf 8}$ & $\overline{\bf 3}$ & $\overline{\bf 6}$
 & $ \overline{\bf 6}$
 & $-4$
		 \\
 \#12 & ${\bf 8}$ & ${\bf 6}$ & ${\bf 3}$
 & ${\bf 3}$
 & $4$
	 \\
 \hline \hline
\end{tabular}
 \caption{Choices of the $SU(3)$ charges of the mediator fields
 and the $SU(3)$ coefficients which appear in the reordering of the
 $SU(3)$ indices to obtain the corresponding effective operator;
 see Eqs.\eqref{eq:SU3-projection} and \eqref{eq:Leff} and the text.
 }
 \label{Tab:SU3-decom}
\end{table}
In the column ``$SU(3)$ coeff.'',
we also list the coefficients appearing in the calculation,
which we call {\it operator projection},
to derive the effective operators from the decompositions.
In order to obtain the effective operators Eqs.~\eqref{eq:O1}-\eqref{eq:O5}
from the decomposition Eq.~\eqref{eq:L-decom-general} where
each Yukawa interaction forms a $SU(3)$ singlet,
we must rearrange the $SU(3)$ indices as
\begin{align}
 \mathscr{L}_{\text{eff}}
 =&
 \wick{1231}{
 [Y_{1}
 \overline{q^{c}}_{I}
 <1\psi^{c} <3S]
 [Y_{2} >1{\overline{\psi^{c}}} q'_{J} <4S'^{\dagger}]
 [Y_{4} \overline{q''^{c}}_{K} <2\psi' >3S^{\dagger}]
 [Y_{3} >2{\overline{\psi'}} \ell >4S']
 }
 \nonumber
 \\
 =&
 Y_{1}Y_{2}Y_{3}Y_{4}
 \times
 \text{$SU(3)$ coeff.}
 \times
 \epsilon^{IJK}
 \overline{q^{c}}_{I}
 q'_{J}
 \overline{q''^{c}}_{K}
 \ell...
 \label{eq:SU3-projection}
 \end{align}
where $I$, $J$, and $K$ are the $SU(3)$ indices
for ${\bf 3}$ representations,
and $\epsilon^{IJK}$ is the total anti-symmetric tensor
to form a singlet with three triplets.
The part omitted from the second line of Eq.~\eqref{eq:SU3-projection},
which is expressed as ``...'',
represents all contents other than the Yukawa couplings ($Y_{1\text{-}4}$),
the coefficient ($SU(3)$ coeff.) brought by
the rearrangement of the $SU(3)$ indices,
and the outer fermion field operators ($q$, $q'$, $q''$, and $\ell$),
such as the propagators of the mediators
and matrices with $SU(2)$ indices.
We have not specified the quark fields at this stage
and rearrange the $SU(3)$ indices by handling
them as {\bf 3} representation field operators in general.
Depending on the decomposition with a specific choice of
the quark fields,
an additional sign can show up in the further rearrangement
of the $SU(3)$ indices, which will be taken into account
after the full information of the decomposition,
with which one can fully specify the ordering of the quark fields.
The sign due to the ordering of the quarks will be given
in Tabs.~\ref{Tab:Summary-O1}-\ref{Tab:Summary-O5}
(as ``$SU(3)$ sign'').
Note that the $SU(2)$ and the Lorentz indices
have not been rearranged at this stage,
and the rearrangement of them will bring other
coefficients and factors.
All the details of the method of decomposition
and operator projection are given in Appendix,
where we demonstrate the derivation of all the coefficients, signs,
and factors, keeping all the indices on the field operators explicitly.

To proceed the operator projection onto the basis operators
Eqs.~\eqref{eq:O1}-\eqref{eq:O5},
we must specify the species of the outer fermion fields,
determine the position of the quark fields on the box diagram,
and identify the $SU(2)$ gauge charges of the mediator fields.
In Tabs.~\ref{Tab:Summary-O1}-\ref{Tab:Summary-O5},
the ways of decomposition are given in the column ``Decom'',
where the given fermion fields correspond to
$(qq')(q''\ell)$ in Fig.~\ref{Fig:Box} and Eq.~\eqref{eq:L-decom-general}.
The electroweak charges of the mediator fields are listed
at the column ``Mediators $SU(2)_{U(1)}$''.
We concentrate on ${\bf 1}$, ${\bf 2}$, and ${\bf 3}$
for the $SU(2)$ representation.
Note that the sign that comes up in the rearrangement
of the $SU(3)$ indices are also given
in the column of ``$SU(3)$ sign''
in Tabs.~\ref{Tab:Summary-O1}-\ref{Tab:Summary-O5},
which cannot be included in Tab.~\ref{Tab:SU3-decom}
because they depend on the ordering of the quark fields
in a decomposition.
We also list the factors and coefficients which come up
in the process of the operator projection after
Eq.~\eqref{eq:SU3-projection}:
the coefficients and signs from the rearrangement of the $SU(2)$ indices
(``$SU(2)$ coeff.''),
the factors and signs from the Fierz transformations (rearrangement of
the Lorentz indices),
and the loop integral factors (``Fierz$\times$Loop factors'').
The functions $I_{4}$ and $J_{4}$ for the loop integrals of
the box diagrams are defined in Appendix. 
In short, once the decomposition (proto-model)
is specified (one from Tab.~\ref{Tab:SU3-decom}
and one from Tabs.~\ref{Tab:Summary-O1}-\ref{Tab:Summary-O5}
are chosen),
the coefficient of the effective operator
is given as
\begin{align}
 \mathscr{L}_{\text{eff}}
 =&
 \text{$SU(3)$ coeff.}
 \times
 \text{$SU(3)$ sign}
 \times
 \text{$SU(2)$ coeff.}
 \nonumber
 \\
 &
 \times
 \text{Fierz factor}
 \times
 \text{Loop factor}
 \times
 Y_{1} Y_{2} Y_{3} Y_{4}
 \nonumber
 \\
 &
 \times
 \text{effective op(s) $\mathcal{O}$
 in Eqs.~\eqref{eq:O1}-\eqref{eq:O5}},
 \label{eq:Leff}
\end{align}
with which, and
also with the help of the nucleon matrix elements
calculated from
lattice~\cite{Yoo:2018fyn,Aoki:2017puj,Aoki:2013yxa,Aoki:2006ib}
and chiral perturbation
theory~\cite{Claudson:1981gh,Chadha:1983sj,Aoki:2008ku},
one can directly calculate the rates of proton decay.
The notations and the derivations of the coefficients, factors,
and signs are given in Appendix.
\begingroup
\squeezetable
\begin{table}[t]
\begin{tabular}{cccccccc}
  \hline \hline
 $\mathcal{O}_{1}$
 & \multicolumn{4}{c}{Mediators $SU(2)_{U(1)}$}
 & $SU(2)$
 & Fierz$\times$Loop
 & $SU(3)$
     \\
 Decom. & $\psi$ & $S$ & $\psi'$ & $S'$ & coeff.
 & factors
 & sign\\
  \hline
  $(du)(QL)$
 & ${\bf 1}_{\alpha}$
     & ${\bf 1}_{\alpha+\frac{1}{3}}$
 & ${\bf 2}_{\alpha+\frac{1}{6}}$
 & ${\bf 1}_{\alpha+\frac{2}{3}}$ & $1$
 & $M_{\psi} M_{\psi'} I_{4}$
 & +
		  \\
  & ${\bf 2}$ & {\bf 2} & {\bf 1} & {\bf 2} & $-1$
	      \\
  & ${\bf 2}$ & {\bf 2} & {\bf 3} & {\bf 2} & $-3$
	      \\
  & ${\bf 3}$ & {\bf 3} & {\bf 2} & {\bf 3} & $3$
 \\
 \hline
  $(ud)(QL)$
 & ${\bf 1}_{\alpha}$
     & ${\bf 1}_{\alpha-\frac{2}{3}}$
 & ${\bf 2}_{\alpha-\frac{5}{6}}$
	     & ${\bf 1}_{\alpha-\frac{1}{3}}$
		 & $1$ & $M_{\psi} M_{\psi'} I_{4}$
 & $-$
		  \\
  & ${\bf 2}$ & {\bf 2} & {\bf 1} & {\bf 2} & $-1$
	      \\
  & ${\bf 2}$ & {\bf 2} & {\bf 3} & {\bf 2} & $-3$
	      \\
  & ${\bf 3}$ & {\bf 3} & {\bf 2} & {\bf 3} & $3$
		  \\
  \hline
  $(dQ)(uL)$
 & ${\bf 1}_{\alpha}$
     & ${\bf 1}_{\alpha+\frac{1}{3}}$
	 & ${\bf 1}_{\alpha-\frac{1}{3}}$
	     & ${\bf 2}_{\alpha+\frac{1}{6}}$
		 & $-1$ & $-\frac{1}{2} J_{4}$
 & $-$
		  \\
  & {\bf 2} & {\bf 2} & {\bf 2} & {\bf 1} & $-1$
		  \\
  & {\bf 2} & {\bf 2} & {\bf 2} & {\bf 3} & $3$
		  \\
  & {\bf 3} & {\bf 3} & {\bf 3} & {\bf 2} & $-3$
 \\
 \hline
  $(Qd)(uL)$ 
 & ${\bf 1}_{\alpha}$
     & ${\bf 2}_{\alpha-\frac{1}{6}}$
	 & ${\bf 2}_{\alpha-\frac{5}{6}}$
	     & ${\bf 1}_{\alpha-\frac{1}{3}}$
		 & $1$ & $\frac{1}{2} J_{4}$
 & $+$
		  \\
  & {\bf 2} & {\bf 1} & {\bf 1} & {\bf 2} & $-1$
		  \\
  & {\bf 2} & {\bf 3} & {\bf 3} & {\bf 2} & $-3$
		  \\
  & {\bf 3} & {\bf 2} & {\bf 2} & {\bf 3} & $3$
 \\
 \hline
  $(uQ)(dL)$
 & ${\bf 1}_{\alpha}$
     & ${\bf 1}_{\alpha-\frac{2}{3}}$
	 & ${\bf 1}_{\alpha-\frac{1}{3}}$
	     & ${\bf 2}_{\alpha+\frac{1}{6}}$
		 & $-1$ & $-\frac{1}{2} J_{4}$
 & $+$
		  \\
  & {\bf 2} & {\bf 2} & {\bf 2} & {\bf 1} & $-1$
		  \\
  & {\bf 2} & {\bf 2} & {\bf 2} & {\bf 3} & $3$
		  \\
  & {\bf 3} & {\bf 3} & {\bf 3} & {\bf 2} & $-3$
		  \\
  \hline
  $(Qu)(dL)$
 & ${\bf 1}_{\alpha}$
     & ${\bf 2}_{\alpha-\frac{1}{6}}$
	 & ${\bf 2}_{\alpha+\frac{1}{6}}$
 & ${\bf 1}_{\alpha+\frac{2}{3}}$ & $1$ & $\frac{1}{2}J_{4}$
 & $-$
		  \\
  & {\bf 2} & {\bf 1} & {\bf 1} & {\bf 2} & $-1$
		  \\
  & {\bf 2} & {\bf 3} & {\bf 3} & {\bf 2} & $-3$
		  \\
  & {\bf 3} & {\bf 2} & {\bf 2} & {\bf 3} & $3$
 \\
  \hline \hline
 \end{tabular}
 \caption{Decomposition (proto-models) of the $d=6$ effective operator
 $\mathcal{O}_{1}=[du][QL]$ which consists of $d$, $u$, $Q$ and $L$.
 The proto-models result in the same effective operator $\mathcal{O}_{1}$
 but with different coefficients, factors, and signs.}
 \label{Tab:Summary-O1}
\end{table}
\endgroup
\begingroup
\squeezetable
\begin{table}[t]
 \begin{tabular}{cccccccc}
  \hline \hline
  $\mathcal{O}_{2}$& \multicolumn{4}{c}{Mediators $SU(2)_{U(1)}$}
  & $SU(2)$
  & Fierz$\times$Loop
  & $SU(3)$
      \\
  Decom. 
      & $\psi$ & $S$ & $\psi'$ & $S'$ & coeff. &
  factors & sign\\
  \hline
  $(Q_{1}Q_{2})(ue)$
  & ${\bf 1}_{\alpha}$
      & ${\bf 2}_{\alpha-\frac{1}{6}}$
	  & ${\bf 2}_{\alpha-\frac{5}{6}}$
  & ${\bf 2}_{\alpha+\frac{1}{6}}$
  & $1$ & $M_{\psi} M_{\psi'} I_{4}$
  & $+$
		  \\
  & {\bf 2} & {\bf 1} & {\bf 1} & {\bf 1} & $1$
		  \\
  & {\bf 2} & {\bf 3} & {\bf 3} & {\bf 3} & $-3$
		  \\
  & {\bf 3} & {\bf 2} & {\bf 2} & {\bf 2} & $3$
		  \\
  \hline
  $(uQ_{2})(Q_{1}e)$
  & ${\bf 1}_{\alpha}$
      & ${\bf 1}_{\alpha-\frac{2}{3}}$
	  & ${\bf 2}_{\alpha-\frac{5}{6}}$
  & ${\bf 2}_{\alpha+\frac{1}{6}}$
  &$1$ & $\frac{1}{2}J_{4}$
  & $-$
		  \\
  & {\bf 2} & {\bf 2} & {\bf 1} & {\bf 1} &$1$
		  \\
  & {\bf 2} & {\bf 2} & {\bf 3} & {\bf 3} &$-3$
		  \\
  & {\bf 3} & {\bf 3} & {\bf 2} & {\bf 2} &$3$
		  \\
  \hline
  $(Q_{1}u)(Q_{2}e)$
  & ${\bf 1}_{\alpha}$
      & ${\bf 2}_{\alpha-\frac{1}{6}}$
	  & ${\bf 1}_{\alpha-\frac{1}{3}}$
  & ${\bf 1}_{\alpha+\frac{2}{3}}$
  & $1$ & $-\frac{1}{2}J_{4}$
  & $-$
		  \\
  & {\bf 2} & {\bf 1} & {\bf 2} & {\bf 2} & $-1$
		  \\
  & {\bf 2} & {\bf 3} & {\bf 2} & {\bf 2} & $-3$
		  \\
  & {\bf 3} & {\bf 2} & {\bf 3} & {\bf 3} & $3$
		  \\
  \hline \hline
 \end{tabular}
 \caption{Decomposition of $\mathcal{O}_{2}$,
 where the ordering of the two quark doublets in the effective operator
 is determined as $\mathcal{O}_{2}\equiv[Q_{1}Q_{2}][ue]$.}
 \label{Tab:Summary-O2}
\end{table}
\endgroup

\begingroup
\squeezetable
\begin{table}[t]
 \hspace*{-0.4cm}
 \begin{tabular}{cccccccc}
  \hline \hline
  $\mathcal{O}_{3,4}$
  & \multicolumn{4}{c}{Mediators $SU(2)_{U(1)}$}
  & $SU(2)$
  & Fierz$\times$Loop
  & $SU(3)$
  \\
  Decom.
  & $\psi$ & $S$ & $\psi'$ & $S'$ &coeff.
		      & factors & sign
		      \\
  \hline 
  $(Q_{1}Q_{2})(Q_{3}L)$
  & ${\bf 1}_{\alpha}$
      & ${\bf 2}_{\alpha-\frac{1}{6}}$
	  & ${\bf 1}_{\alpha-\frac{1}{3}}$
  & ${\bf 2}_{\alpha+\frac{1}{6}}$ &
  $-\frac{1}{2} \mathcal{O}_{3} + \frac{1}{2} \mathcal{O}_{4}$
  & $M_{\psi} M_{\psi'} I_{4}$
  & $+$
		      \\
  & {\bf 1} & {\bf 2} & {\bf 3} & {\bf 2}
		& $ -\frac{3}{2} \mathcal{O}_{3} - \frac{1}{2}
		      \mathcal{O}_{4}$
		      \\
  & {\bf 2} & {\bf 1} & {\bf 2} & {\bf 1}
		  & $ \mathcal{O}_{3} $
		      \\
  & {\bf 2} & {\bf 1} & {\bf 2} & {\bf 3}
		  & $ \mathcal{O}_{4} $
		      \\
  & {\bf 2} & {\bf 3} & {\bf 2} & {\bf 1}
		  & $ -\mathcal{O}_{4} $
		      \\
  & {\bf 2} & {\bf 3} & {\bf 2} & {\bf 3}
		  & $ -3 \mathcal{O}_{3} + 2 \mathcal{O}_{4} $
		      \\
  & {\bf 3} & {\bf 2} & {\bf 1} & {\bf 2}
		  & $ - \frac{3}{2} \mathcal{O}_{3} - \frac{1}{2}
		      \mathcal{O}_{4}$
		      \\
  & {\bf 3} & {\bf 2} & {\bf 3} & {\bf 2}
  & $ -\frac{9}{2} \mathcal{O}_{3}
  + \frac{1}{2} \mathcal{O}_{4}$
		      \\
  \hline \hline
 \end{tabular}
 \caption{Decompositions of the effective operators with
 three $Q$s and a $L$.
 Each model results in a different combination of $\mathcal{O}_{3}$
 and $\mathcal{O}_{4}$.
 Note that the ordering of the three $Q$s in the effective operators
 are fixed as $\mathcal{O}_{3/4}$ $\equiv [Q_{1}Q_{2}]_{{\bf 1}/{\bf
 3}}[Q_{3}L]_{{\bf 1}/{\bf 3}}$.}
 \label{Tab:Summary-O3O4}
 \end{table}
\endgroup
\begingroup
\squeezetable
\begin{table}[t]
 \begin{tabular}{ccccccc}
  \hline \hline
  $\mathcal{O}_{5}$ &
     \multicolumn{4}{c}{Mediators $SU(2)_{U(1)}$}
  & $SU(2)$
  & Fierz$\times$Loop factors
  \\
  Decom. & $\psi$ & $S$ & $\psi'$ & $S'$ & coeff.
  & $\times SU(3)$ sign 
  \\
  \hline
  $(du_{1})(u_{2}e)$
  &
      ${\bf 1}_{\alpha}$
      & ${\bf 1}_{\alpha+\frac{1}{3}}$
	  & ${\bf 1}_{\alpha-\frac{1}{3}}$
  & ${\bf 1}_{\alpha+\frac{2}{3}}$ & $1$ &
  $M_{\psi}M_{\psi'}I_{4}\mathcal{O}_{5}$
  \\
  &
      {\bf 2} & {\bf 2} & {\bf 2} & {\bf 2} & $2$
		  \\
  &
      {\bf 3} & {\bf 3} & {\bf 3} & {\bf 3} & $3$
  \\
    \hline
  $(u_{1} d)(u_{2} e)$
  &
      ${\bf 1}_{\alpha}$
      & ${\bf 1}_{\alpha-\frac{2}{3}}$
	  & ${\bf 1}_{\alpha-\frac{4}{3}}$
  & ${\bf 1}_{\alpha-\frac{1}{3}}$ & $1$ &
  $-M_{\psi}M_{\psi'}I_{4}\mathcal{O}_{5}$
  \\
  &
      {\bf 2} & {\bf 2} & {\bf 2} & {\bf 2} & $2$
		  \\
  &
      {\bf 3} & {\bf 3} & {\bf 3} & {\bf 3} & $3$
  \\
    \hline
  $(u_{2}u_{1})(de)$
  &
      ${\bf 1}_{\alpha}$
      & ${\bf 1}_{\alpha-\frac{2}{3}}$
	  & ${\bf 1}_{\alpha-\frac{1}{3}}$ &
  ${\bf 1}_{\alpha+\frac{2}{3}}$ & $1$
  & $M_{\psi}M_{\psi'}I_{4} \left[ \mathcal{O}_{5} - \mathcal{O}_{5}'\right]$
  		  \\
  &
      {\bf 2} & {\bf 2} & {\bf 2} & {\bf 2} & $2$
		      &

		  \\
  &
      {\bf 3} & {\bf 3} & {\bf 3} & {\bf 3} & $3$
		  \\
  \hline \hline 
 \end{tabular}
 \caption{Decompositions of $\mathcal{O}_{5}$.
 Here the ordering of the two $u$'s in the basis operators
 is determined as $\mathcal{O}_{5}\equiv[du_{1}][u_{2}e]$
 and $\mathcal{O}_{5}'\equiv[du_{2}][u_{1}e]$.}
 \label{Tab:Summary-O5}
\end{table}
\endgroup

Here we show an example to demonstrate
how to use the information of the tables.
A famous dimension-five contribution to proton decay
in SUSY-GUT models
is found by taking \#1 from Tab.~\ref{Tab:SU3-decom}
and the decomposition of the seventh row in Tab.~\ref{Tab:Summary-O3O4}
with $\alpha=0$ for $U(1)$ hypercharge.
The mediators are identified with the SUSY particles as
\begin{gather}
 \psi({\bf 1},{\bf 3})_{0} = \widetilde{W},
 \quad
 S(\overline{\bf 3}, {\bf 2})_{-1/6} = \widetilde{Q}^{*}
 \nonumber
 \\
 \psi'({\bf 3},{\bf 1})_{-1/3} = \widetilde{h}_{c},
 \quad
 S'({\bf 3},{\bf 2})_{+1/6} = \widetilde{Q},
 \label{eq:model-for-d5SUSYGUT}
\end{gather}
where $\widetilde{W}$ is the wino,
$\widetilde{h}_{c}$ is the coloured higgsino,
and $\widetilde{Q}$ is the squark doublet.
Using the information listed in the tables,
we can reproduce the coefficient of
the effective operator Eq.~\eqref{eq:Leff} as
\begin{align}
 \mathscr{L}_{\text{eff}}
 =&
 \overbrace{
 (-1)
 }^{\text{SU(3) coeff.}}
 \times
 \overbrace{
 (+)
 }^{\text{$SU(3)$ sign}}
 \times
 \overbrace{
 \left[
 -\frac{3}{2}
 \mathcal{O}_{3}
 -
 \frac{1}{2}
 \mathcal{O}_{4}
 \right]
 }^{\text{$SU(2)$ coeff. and $\mathcal{O}$}}
 \nonumber
 \\
 &
 \times
 \overbrace{
 M_{\psi} M_{\psi'} I_{4}
 }^{\text{Fierz$\times$Loop factors}}
 \times
 Y_{1} Y_{2} Y_{3} Y_{4}
 \nonumber
 \\
 =&
 \frac{1}{2}
 Y_{1}Y_{2}Y_{3}Y_{4} M_{\widetilde{W}}
 M_{\widetilde{h}_{c}}
 I_{4}
 \left[
 3
 \mathcal{O}_{3}
 +
 \mathcal{O}_{4}
 \right].
\end{align}
%
Note that 
$Y_{1}$ and $Y_{2}$ are given by the gauge coupling
in $SU(5)$ SUSY-GUT models.
The coupling $Y_{3}$ is identified with the coupling for
the Yukawa interaction of
${\bf 10} \cdot \overline{\bf 5} \cdot  H(\overline{\bf 5})$,
and $Y_{4}$ is that for ${\bf 10} \cdot {\bf 10} \cdot H({\bf 5})$,
where
${\bf 10}$ and $\overline{\bf 5}$ are
the matter superfields
and
$H(\overline{\bf 5})$ and $H({\bf 5})$
are the Higgs superfields.
Taking the decomposition of the first row in
Tab.~\ref{Tab:Summary-O3O4},
one can find the same diagram but with 
a bino $\widetilde{B}({\bf 1},{\bf 1})_{0}$
instead of the wino $\widetilde{W}$.

\section{Models and phenomenology}

In this section we discuss the phenomenology of the different 1-loop
models presented above. We will start with a brief discussion of the
different model classes and an overview of commonalities that all
these 1-loop models share.  We then discuss two example models in some
more detail. 

\subsection{General discussion\label{sect:3.1}}

Our results listed in Tabs.~\ref{Tab:SU3-decom}-\ref{Tab:Summary-O5}
summarize the possible particle content that
allows to construct models with 1-loop induced proton decay, we call
this the ``proto-models''.
However, not all allowed choices of quantum numbers will automatically
result in models, in which the 1-loop contribution to proton decay will
be the dominant one.
To see this in a simple example, consider decomposition \#1 from
Tab.~\ref{Tab:SU3-decom} and the decomposition of the third row in
Tab.~\ref{Tab:Summary-O3O4} with $\alpha=1/2$. 
In this case $S$ is identified with $S(\overline{\bf 3}, {\bf
1})_{1/3}$.
The quantum numbers of this scalar allow to write down the following two
interactions with standard model fermions:
$QQS^{\dagger}$ and $QLS$.
The product of these interactions,
after integrating out $S$, generate ${\mathcal O}_3$
{\em at tree-level}.\footnote{%
$S(\overline{\bf 3}, {\bf 1})_{1/3}$ is not the only choice,
that will lead to tree-level proton decay. The same argument applies
to $S(\overline{\bf 3}, {\bf 3})_{1/3}$ and $S(\overline{\bf 3},
{\bf 1})_{4/3}$.}
Thus, unless there is a strong hierarchy between the different
Yukawa interactions, one expects that the tree-level contribution
dominates the decay rate.
One can eliminate such an unwanted hierarchy in couplings using
additional symmetries. The simplest possibility is to just assign
the particles running in the loop to be odd under a $Z_2$,
while all the standard model particles are even.
We classify models, which need such an additional symmetry
to avoid unwanted tree-level proton decay, as class-I models.
We discuss one example model from this class in Section \ref{sect:modI}.

In addition, there exist choices of quantum numbers, for which
tree-level 2-body decays are not allowed, but higher multiplicity
final states are generated at tree-level together with the 1-loop
diagrams.
For example, if $S$ and $S'$ are chosen to be $S({\bf 3},
{\bf 2})_{1/6}$ and $S'({\bf 1}, {\bf 2})_{1/2}$,
which have the same charges as a scalar leptoquark
and the SM Higgs field,
the corresponding models will produce 3-body proton decays,
such as $p\to \pi^+ \pi^+ e^-$ via an effective $d=9$ operator.
In such cases, one expects in general that the $d=6$ 1-loop operator
dominates over the $d=9$ tree-level operator for typical scales
$\Lambda \gsim 1$ TeV, see Eq.~\eqref{eq:pdec}.
However, if there is some hierarchy in the Yukawa
couplings, $(Y_3Y_4) \ll (Y_1Y_2)$, one can arrange the 3-body decays
to dominate over the 2-body ones and one needs again a symmetry to
assure that the loop dominates over the tree-level contribution.
Our first example model is exactly of this type, see Section
\ref{sect:modI}.

Finally, there are choices, where the particle content of the 1-loop
model is such that tree-level proton decay can occur only at $d=12$
and higher (usually leading to proton decay with 5-body final
states). In these cases the 1-loop $d=6$ decay will win over the
tree-level decays for all practical choices of model parameters. We
consider such models interesting and define these models as
``class-II'' models, since no symmetry is required to make the 1-loop
$d=6$ decays dominant. We will discuss one concrete example model in
Section \ref{sect:modII}.

Obviously, the main difference between models in class-I and class-II 
is that in class-I the lightest particle will be absolutely stable.
This opens up the possibility to connect proton decay to dark 
matter, but requires an electrically neutral particle in the loop.
We will come back to a more detailed discussion of this point in 
Section \ref{sect:modI}.

Let us now turn to a rough estimate of the proton decay half-life.
Using results from lattice QCD
calculations~\cite{Yoo:2018fyn,Aoki:2017puj,Aoki:2013yxa,Aoki:2006ib}
and chiral perturbation
theory~\cite{Claudson:1981gh,Chadha:1983sj,Aoki:2008ku},
the two-body proton decay half-life can be calculated as:\footnote{%
The simple estimate, Eq.~\eqref{eq:pdec} would give 
${\bar Y}$ roughly a factor 2.5 smaller.}
\begin{align}
 \tau
 \simeq&
 \frac{1}{\frac{m_{p}}{32\pi}
 \left[
 1 - \frac{m_{\text{meson}}^{2}}{m_{p}^{2}}
 \right]^{2}
 \left|
 W \frac{Y_{1}Y_{2}Y_{3}Y_{4}}{16 \pi^{2} \cdot 6 \cdot M^{2}}
 \right|^{2}}
 \nonumber
 \\
 \sim &
 10^{34}\text{[yrs]}
 \left[
 \frac{M}{1\text{TeV}}
 \right]^{4}
 \left[
 \frac{3 \times 10^{-6}}{{\bar Y}}
 \right]^{8}.
 \label{eq:lifetime-estimation}
\end{align}
Note that half-live estimates for different operators,
${\cal O}_1$-${\cal O}_5$,
differ slightly due to the different possible final states.
For the numerical estimate we use the charged pion mass for
the mass $m_{\text{meson}}$ of the daughter meson.
$W$ is the corresponding hadronic matrix element,
$W \equiv \langle \text{meson}|(qq)q | p \rangle = -0.181$ [GeV$^{2}$];
the numerical value has recently been calculated in
Ref.~\cite{Aoki:2017puj}.
The factor $1/6$ in the first equation above is due to the loop
integral $I_{4}$, in the limit of equal masses.
We have defined the mean coupling ${\bar Y}=(Y_{1}Y_{2}Y_{3}Y_{4})^{1/4}$,
since proton decay is sensitive only to this product and used
a mass scale of $1$ TeV, since we are interested in possible
LHC phenomenology of these 1-loop models.

With couplings of the order of Eq.~\eqref{eq:lifetime-estimation}, the
particles in the 1-loop diagrams can be rather long-lived.
Depending on the choices of parameters, i.e. Yukawa couplings and mass
hierarchies of the new particles, decay lengths can vary from
unmeasurably short to many meters.
The collider phenomenology of long-lived particles has recently
attracted a lot of attention in the literature, see for example
Refs.~\cite{Helo:2013esa,delaPuente:2015vja,Liu:2018wte,Lee:2018pag,Belanger:2018sti,Cottin:2018kmq,Cottin:2018nms,Cottin:2019drg}.
There are also plans for several future experiments, dedicated to the
search for ultra long-lived particles, see for example
Refs.~\cite{Chou:2016lxi,Gligorov:2018vkc,Dercks:2018wum,Helo:2018qej,Curtin:2018mvb}.
For the current status of searches for long-lived particles at the LHC,
see
Refs.~\cite{Sirunyan:2018vlw,Aaboud:2018aqj,Sirunyan:2018njd,Aaboud:2018hdl,Aaboud:2018jbr,Sirunyan:2018pwn,Sirunyan:2018ldc,Sirunyan:2018vjp,Sirunyan:2017sbs,Aaboud:2017mpt,Sirunyan:2017jdo,Aaboud:2017iio,Khachatryan:2016unx,Aaboud:2016dgf,Aad:2015rba,CMS:2014hka,CMS:2014wda,Chatrchyan:2013oca}.
We will come back to a more detailed discussion of the LHC
phenomenology of our 1-loop models in Sections \ref{sect:modI} and
\ref{sect:modII} below.

\subsection{Model I\label{sect:modI}}

\begin{figure}[t]
 \unitlength=1cm
 \begin{picture}(5.5,3.5)
 \put(0,0){\includegraphics[width=5.3cm]{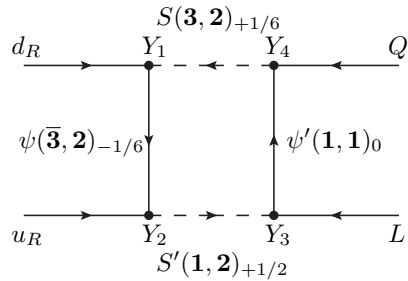}}
   \end{picture}
 \caption{An example, Model-I, for a one-loop decomposition of the
   proton decay operator $\mathcal{O}_{1}$. This model relates proton
   decay to neutrino masses, the dark matter, and possibly a
   long-lived coloured particle at the LHC, compare with
   Fig.~\ref{Fig:Pheno-Model-Nr1}.}
 \label{Fig:Model-Nr1}
\end{figure}

Here we discuss one example model, corresponding to the choices \#5 in
Tab.~\ref{Tab:SU3-decom} for colour, the second row in
Tab.~\ref{Tab:Summary-O1} for the decomposition of $\mathcal{O}_{1}$,
and the parameter $\alpha$ for the electroweak $U(1)$ hypercharge to
$-1/6$.
The SM charges of the mediator fields are then determined as
\begin{align}
 \psi(\overline{\bf 3},{\bf 2})_{-1/6},
 \quad
 S({\bf 3},{\bf 2})_{+1/6},
 \nonumber
 \\
 \psi'({\bf 1},{\bf 1})_{0},
 \quad
 S'({\bf 1},{\bf 2})_{+1/2}.
\end{align}
Note that the scalar mediator field $S'$ has the same charges
as the SM Higgs field $H$. The 1-loop diagram for proton 
decay is shown in Fig.~\ref{Fig:Model-Nr1}. 

As discussed above, the model allows for a $d=9$ 3-body decay of the
proton, unless an additional symmetry is introduced.
This can be seen easily as follows:
Cutting the diagram in the scalar lines, one obtains the effective
operator $d_R u_R S S'^{\dagger}$.
Including the decays of the scalars produces then
a $\Delta(B-L)=2$ $d=9$ operator:
$(d_R u_R)(\overline{L} d_R)(\overline{u_R} Q)$.
It is easy to forbid this
operator, simply postulating $B-L$ conservation, since the original
$d=6$ operator has $\Delta(B-L)=0$.
More interesting phenomenologically, however,
is to assign a new $Z_2$ to the model,
under which all loop particles are odd, while the SM fermions are even.
In this case, unless the $Z_2$ is spontaneously broken, the
lightest of the particles in the loop is absolutely stable and 
can be therefore a candidate for the dark matter. In the following, 
we will discuss this variant of the model. 

Combining the information listed in Tabs.~\ref{Tab:SU3-decom} and
\ref{Tab:Summary-O1}, one can find the coefficient of the effective
interaction of proton decay processes as
\begin{align}
 \mathcal{C}_{1}
 =
 -M_{\psi} M_{\psi'} I_{4}
 Y_{1}Y_{2}Y_{3}Y_{4},
 \label{eq:C1-Model-Nr1}
\end{align}
with which the effective Lagrangian Eq.~\eqref{eq:Leff} is given as
$\mathscr{L}_{\text{eff}} \equiv \mathcal{C}_{1} \mathcal{O}_{1}$.
The effective operator $\mathcal{O}_{1}$ causes decay of a proton in
two modes, and the rates are calculated with the coefficient
Eq.~\eqref{eq:C1-Model-Nr1} as\footnote{In this estimate we do not
take into account the effect of the renormalization group
running~\cite{Abbott:1980zj,Alonso:2014zka} of the operators, and
use the coefficient at the scale of the proton mass.
This is sufficient for our rough estimates.}
\begin{align}
 \Gamma (p \rightarrow \pi^{+} \bar{\nu}_{e}/\pi^{0} e^{+}) =&
  \frac{m_{p}}{32 \pi}
 \left[
 1 - \frac{m_{\pi^{+/0}}^{2}}{m_{p}^{2}}
 \right]^{2}
 \left|
 W_{0}
 \mathcal{C}_{1}
 \right|^{2},
 \label{eq:Pdecayrate-formula}
\end{align}
where the hadronic matrix elements $W_{0}$ are found in
Ref.~\cite{Aoki:2017puj}: $W_{0} = -0.186 (-0.131)$ [GeV$^{2}$] for
the $\pi^{+}$ ($\pi^{0}$) mode.
All decompositions in Tab.~\ref{Tab:Summary-O1}, which result in the operator
$\mathcal{O}_{1}$, predict roughly the same size of the rates for both decay 
modes.
Therefore, if it turns out that the rates of the two modes are
very different, models based on $\mathcal{O}_{1}$ will be disfavored.
As we have already seen in Eq.~\eqref{eq:lifetime-estimation},  the 
mean of the couplings ${\bar Y}=(Y_{1}Y_{2}Y_{3}Y_{4})^{1/4}$ should be 
order few ${\cal O}(10^{-6})$  for masses accessible at the LHC.

The interaction
$Y_{3} \overline{\psi'}({\bf 1},{\bf 1})_{0} L S'({\bf 1},{\bf 2})_{+1/2}$
in Fig.~\ref{Fig:Model-Nr1} 
can be identified with the corresponding interaction that appears in
the scotogenic model~\cite{Ma:2006km},
since $\psi'({\bf 1},{\bf 1})_{0}$ can be interpreted as a $\nu_R$.
Note that Model-I is not the
only decomposition that contains such an interaction.
Requiring the fields ($\psi'$ and $S'$) relevant for the radiative
neutrino mass generation to be colour singlets, we have only one choice
left for the assignment of the colour charges, which is \#5 in
Tab.~\ref{Tab:SU3-decom}.
Assuming the $\psi'$ to be a singlet under the electroweak $SU(2)$
as in the original scotogenic model, we have the choices
\#2, 6, 9, 14, 17, 22 in Tab.~\ref{Tab:Summary-O1} for
$\mathcal{O}_{1}$ and \#1 and 7 in Tab.~\ref{Tab:Summary-O3O4} for
$\mathcal{O}_{3,4}$.
In total, we have eight possibilities for loop-induced proton decays
which can accommodate neutrino masses and dark matter with
the scotogenic-type realization.\footnote{%
The scotogenic type diagram for neutrino masses can be
drawn with the Majorana fermion ($\psi'$) in the adjoint
representations under the SM gauge symmetries.  If we relax the
requirements to include the adjoint representations, we have more
possibilities: \#6, 7, and 8 in Tab.~\ref{Tab:SU3-decom} for colour,
and \#3, 7, 12, 15, 20, and 23 in Tab.~\ref{Tab:Summary-O1} for
$\mathcal{O}_{1}$ and \#2, and 8 in Tab.~\ref{Tab:Summary-O3O4} for
$\mathcal{O}_{3,4}$.}

The phenomenology of the scotogenic model has been studied in many
papers, see for example
Refs.~\cite{Kubo:2006yx,Boehm:2006mi,Sierra:2008wj,Suematsu:2009ww,%
Gelmini:2009xd,Schmidt:2012yg,Restrepo:2013aga,Racker:2013lua,%
Toma:2013zsa,Molinaro:2014lfa,Vicente:2014wga,Hessler:2016kwm,%
Hagedorn:2018spx}.
We will therefore only briefly summarize the most important aspects of
its phenomenology here and comment on the differences between our
Model-I and the original scotogenic model.

\begin{figure}[t]
 \unitlength=1cm
\begin{picture}(8,7)
\put(-0.5,0.5){\includegraphics[width=9cm]{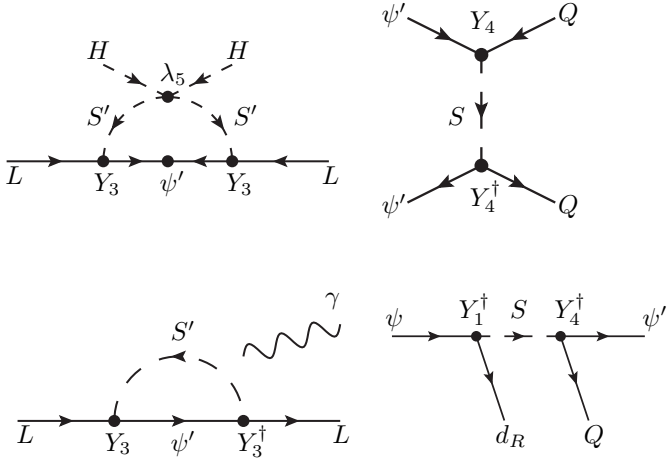}}
\end{picture}
 \caption{Phenomenology of Model-I: Majorana masses for neutrinos
   (upper left), example of a DM annihilation process for the freeze-out
   scenario (upper right), a charged lepton flavour violating process
   (lower left), and a decay chain of the coloured mediator field
   $\psi$ at the LHC (lower right).  }
 \label{Fig:Pheno-Model-Nr1}
\end{figure}

To generate a Majorana mass term for neutrinos from the 1-loop
diagram (upper-left in Fig.~\ref{Fig:Pheno-Model-Nr1}) one introduces
a scalar quartic interaction between the SM Higgs and the 
new scalar~\cite{Ma:2006km}:
\begin{align}
 \mathscr{L}
 \supset
 \lambda_5
 (S'^{\dagger} H)(S'^{\dagger} H)
 +
 {\rm H.c.}
 \label{eq:L-scoto-H-Sdash-mixing}
\end{align}
The flavour structure of the Majorana mass term for neutrinos
in the scotogenic model can be expressed as
\begin{align}
 (m_{\nu})^{\alpha \beta} =
 \sum_{i}
 {(Y_{3}^{\sf T})^{\alpha}}_{i}
 \Lambda_{i}
 {(Y_{3})_{i}}^{\beta}.
\end{align}
which shows that at least two $\psi'$ are necessary to reproduce
the two mass squared differences measured by neutrino oscillation
data.
One can find the loop integral $\Lambda_{i}$ in Ref.~\cite{Ma:2006km},
which is 
\begin{align}
 \Lambda_{i}
 \equiv&
 \frac{M_{\psi'_{i}}}{16 \pi^{2}}
 \left[
 f
 (
 M_{\psi'_{i}}^{2}/M_{\text{Re}S'^{0}}^{2}
 )
 -
 f
 (
 M_{\psi'_{i}}^{2}/M_{\text{Im}S'^{0}}^{2}
 )
 \right]
 \label{eq:scoto-Lambda}
\end{align}
where the function $f(x)$ is defined as $f(x) = -\ln t/(1 - x)$.
The splitting between the mass $M_{\text{Re}S'^{0}}$ of the real part
of the neutral component of $S'$ and that $M_{\text{Im}S'^{0}}$ of the
imaginary part is given by the scalar mixing term
Eq.~\eqref{eq:L-scoto-H-Sdash-mixing} as $M_{\text{Re}S'^{0}}^{2} -
M_{\text{Im}S'^{0}}^{2} = 2 \lambda_{5} \langle H^{0} \rangle^{2}$.
It is clear from Eq.~\eqref{eq:scoto-Lambda} and also the diagram in
Fig.~\ref{Fig:Pheno-Model-Nr1} that the loop integral vanishes in the
limit where the mass splitting, which is proportional to the scalar
mixing, goes to zero.
In short, the size of the neutrino masses is controlled by the scalar
mixing coupling $\lambda_5$, the Yukawa interaction $Y_{3}$, and
the masses of the mediators.
Here we are interested in the phenomenology of the mediators with
masses of the TeV scale.
Setting the mediator masses to roughly a TeV, one
finds that either $Y_{3}$ should be small, say of the order of ${\cal
  O}(10^{-5})$ for $\lambda_5$ order ${\cal O}(1)$, or $\lambda_5$
should be order ${\cal O}(10^{-10}\text{-}10^{-8})$ for Yukawas 
order $0.1$-$1$ to obtain $m_{\nu} \sim \mathcal{O}(0.1)$ eV.

The $Y_{3}$ interaction also mediates charged lepton flavour violating
(cLFV) processes.
Studies with a parameter scan, e.g.,
Refs.~\cite{Toma:2013zsa,Vicente:2014wga}, conclude that
the $\ell_{\alpha} \rightarrow \ell_{\beta} \gamma$ processes
{\em currently} places the most stringent constraints on the model
parameters in wide area of the parameter space.
A general formula for the rate of this cLFV process has been presented in
Ref.~\cite{Lavoura:2003xp}.
It can be written as
\begin{align}
 \Gamma (\ell_{\alpha} \rightarrow \ell_{\beta} \gamma)
 =
 \frac{e^{2} m_{\ell_{\alpha}}^{5}}{16\pi}
 \left|
 \sum_{i}
 (Y_{3}^{\dagger}{)_{\beta}}^{i}
 (Y_{3}{)_{i}}^{\alpha}
 \left[
 -\overline{c}
 +
 \frac{3}{2}\overline{d}
 \right]_{i}
 \right|^{2}.
 \label{eq:cLFV}
\end{align}
The loop integral factor in Eq.~\eqref{eq:cLFV} is given as
\begin{align}
 \left[
 -\overline{c}
 +
 \frac{3}{2} \overline{d}
 \right]_{i}
 =
 \frac{{\rm i}}{16 \pi^{2}}
 \frac{1}{M_{S'^{+}}^{2}}
 \left[
 \frac{2 t_{i}^{2} + 5t_{i} -1}{12 (t_{i}-1)^{3}}
 -
 \frac{t_{i}^{2} \ln t_{i}}{2 (t_{i}-1)^{4}}
 \right]
\end{align}
with $t_{i} \equiv M_{\psi'_{i}}^{2}/M_{S'^{+}}^{2}$.
The non-observation of
the $\mu \rightarrow e \gamma$ process~\cite{TheMEG:2016wtm}
suggests ${(Y_{3})_{i}}^{\alpha\in\{e,\mu\}} \lesssim \mathcal{O}(0.01\text{-}1)$
for mediators with masses of $\mathcal{O}(1)$ TeV~\cite{Vicente:2014wga}.
For future prospects of the experimental bounds to the cLFV
processes are summarized in e.g., Ref.~\cite{Calibbi:2017uvl}.

We now turn to a brief discussion of dark matter.
There are two possible candidates in Model-I.
The scalar $S'({\bf 1},{\bf 2})_{+1/2}$ can be identified with
the inert doublet, discussed many times in the literature.
For inert doublet DM see, for example
Refs.~\cite{LopezHonorez:2006gr,Arhrib:2013ela}.
The second candidate is the neutral fermion.
For a detailed study of singlet fermion  DM in the scotogenic 
model see, for example Refs.~\cite{Vicente:2014wga,Hagedorn:2018spx}.

Suppose the lightest Majorana fermion $\psi'_{1}$ is the DM field and
was thermally produced and frozen out in the early Universe.
In our model, there are two pair-annihilation modes for $\psi'_{1}$,
which are $\psi'_{1} \psi'_{1} \rightarrow Q{\bar Q}$ and $L{\bar L}$.
The $L{\bar L}$ mode, mediated by the $Y_{3}$ interaction, is the only
mode in the original scotogenic model.
For this diagram to be efficient enough, $Y_{3}$ should be large,
which produces a mild tension between upper limits from cLFV and
the minimal $Y_{3}$ required to reproduce the correct relic
density~\cite{Vicente:2014wga}.\footnote{%
It is possible to solve this problem of the overabundance of
DM field in scotogenic models in regions of parameter space
where co-annihilation processes are sizeable, see for example
Refs.~\cite{Vicente:2014wga,Hagedorn:2018spx}.}
However, the model we are discussing here also has the interaction
$Y_{4}$ of the DM field $\psi'$ with $Q$ and can annihilate through
the $\psi'\psi'\rightarrow Q{\bar Q}$ channel, see
Fig.~\ref{Fig:Pheno-Model-Nr1}.
The cross section for this pair-annihilation process can
be roughly estimated as
\begin{align}
 &\sum_{q=u,d}
 \langle
 \sigma(\psi' \psi' \rightarrow q \bar{q})
 v
 \rangle
 \sim
 2
 \times
 \pi
 \left[
 \frac{|Y_{4}|^{2}}{4\pi}
 \right]^{2}
 \frac{1}{M^{2}}
 \times
 \frac{T}{M}
 \nonumber
 \\
 &=
 2 \cdot 10^{-26}
 \text{[cm$^{3}$/s]}
 \left[
 \frac{Y_{4}}{1.0}
 \right]^{4}
 \left[
 \frac{\text{TeV}}{M}
 \right]^{2}
\end{align}
where $T$ is the freeze-out temperature.
The suppression factor $T/M \sim 1/20$ comes from the fact that the
annihilation amplitude is $p$-wave, since the initial state consists
of two Majorana fermions, cf. e.g., Ref.~\cite{Sigl:2017wfx}.
Note that, cross sections of order $2 \cdot 10^{-26}$ [cm$^{3}$/s]
will reproduce the correct relic density.
Again, as in the case of $Y_3$, much smaller values of this coupling
would be sufficient, if $S({\bf 3},{\bf 2})_{+1/6}$ is not much
heavier than $\psi'$, such that co-annihilation effects become important.


Finally, we will discuss the LHC phenomenology of Model-I.
We will concentrate on the coloured states $S({\bf 3},{\bf 2})_{+1/6}$
and $\psi(\overline{\bf 3},{\bf 2})_{-1/6}$.
Let us first consider $S$.
This scalar will decay to a 2-body final state of jets ($j$)
plus missing energy ($\esl$).
The decay will be prompt, unless $Y_4$ is tiny, say $Y_4 \ll 10^{-7}$.
Thus, limits from standard SUSY searches apply.
For example, CMS has searched for scalar quarks decaying promptly
to jets plus missing energy \cite{Sirunyan:2017cwe}.
The limits for one generation of squarks reach up to 1 TeV
for neutralino mass of $m_{\tilde{\chi}^{0}} \sim 100$ GeV and weaken to roughly
600 GeV for $m_{\tilde{\chi}^{0}} \sim 400$ GeV~\cite{Sirunyan:2017cwe}.
Similar numbers can be found in ATLAS searches,
for example~\cite{Aaboud:2017vwy}.

The possible decay chains for the coloured fermion 
$\psi^{c}({\bf 3},{\bf 2})_{+1/6}$
lead to final states of either $jj\esl$
or $jl^{\pm}\esl$.
Thus, even though $\psi$ resembles
a vector-like quark (VLQ) from its quantum numbers, standard
VLQ searches do not apply to this state.
(For a summary of CMS searches for VLQs see, for example
Ref.~\cite{Beauceron:2651966}.)
On the other hand, pair production of $\psi$ will
lead to final states that resemble again those for SUSY searches for
squarks and gluinos.
However, which of the LHC searches can be used to constrain
the $\psi$ depends on whether its decays are prompt or not.
This in turn depends on the mass hierarchy
of the particles in the loop, see Fig.~\ref{Fig:Pheno-Model-Nr1}.
If the mass $M_\psi$ of $\psi$  
is larger than either of the masses of the scalars 
$S({\bf 3},{\bf 2})_{+1/6}$ or $S'({\bf 1},{\bf 2})_{+1/2}$, 
the decays of $\psi$ are 2-body and likely prompt,
unless again the corresponding Yukawa is 
considerably smaller than the estimate for ${\bar Y}$ of 
${\cal O}(10^{-6})$ discussed above from proton decay 
sensitivities.
ATLAS gives lower 
limits on the gluino mass in simplified SUSY models 
of order $m_{\tilde{g}}\gsim 2$ TeV~\cite{Aaboud:2017vwy}.
However, the limits on $\psi$ will be weaker, since 
(a) the cross section for a colour triplet is smaller 
than for the gluino (octet) and
(b) the $\psi$ can also decay
to $jl^{\pm}\esl$ with an unknown branching ratio,
so this 0-lepton search \cite{Aaboud:2017vwy} does 
not always directly apply.
Masses $M_{\psi}$ below 1 TeV 
will, however, always be excluded since ATLAS leptoquark 
searches \cite{Aaboud:2016qeg} can be combined with the 
SUSY search \cite{Aaboud:2017vwy}, as long as the decays 
of $\psi$ are prompt.

Assume now that $M_\psi$
is smaller than the mass $M_{S}$
of the scalar $S({\bf 3},{\bf 2})_{+1/6}$.
The three-body decay rate of $\psi$ to $jj\esl$ can then be estimated as
\begin{equation}\label{eq:3bdecay}
\Gamma(\psi(\overline{\bf 3},{\bf 2})_{-1/6} \to jj\esl) 
 \simeq
 \frac{\left| Y_1Y_4 \right|^2}{512 \pi^3}
 \left[
  \frac{M_\psi}{M_S}
 \right]^4
 M_\psi .
\end{equation}
Note that the rate for the three-body final state $jl^{\pm}\esl$ is
given by the same expression, simply replacing $|Y_1Y_4|$ with $|Y_2Y_3|$
and taking $M_S$ as the mass of $S'({\bf 1},{\bf 2})_{+1/2}$.
From Eq.~\eqref{eq:3bdecay} one can estimate that for $M_\psi \gsim
M_S \simeq 1$ TeV decay lengths will become larger than the order of
millimetre for Yukawas smaller than $10^{-3}$.
For Yukawas as small as $10^{-6}$, see Eq.~\eqref{eq:lifetime-estimation},
life-times exceed already 10 seconds.
Thus, the $\psi$ will hadronize before decaying.
ATLAS studied constraints on long-lived coloured
particles~\cite{Aaboud:2019trc},
again in the context of a supersymmetric model.
From Figs.~9 and 11 in Ref.~\cite{Aaboud:2019trc}
one can estimate that $\psi$ should
be heavier than $M_\psi \gsim$ 1.8-1.9 TeV for
$c\tau=$3-10 m.
From Ref.~\cite{Heinrich:2018pkj} one can estimate that similar numbers will
apply also for quasi-stable $\psi$. 

In summary, Model-I allows to connect proton decay, dark matter and
neutrino masses.
If the masses of the loop particles are of order of ${\cal O}
(1\text{-}2)$ TeV,
one can have also a wide range of interesting
signals at the LHC.
We have discussed a few possible search strategies
for the LHC for the coloured particles in this model.

\subsection{Model II}
\label{sect:modII}

\begin{figure}[t]
 \unitlength=1cm
 \begin{picture}(5.5,3.5)
  \put(0,0){\includegraphics[width=5.3cm]{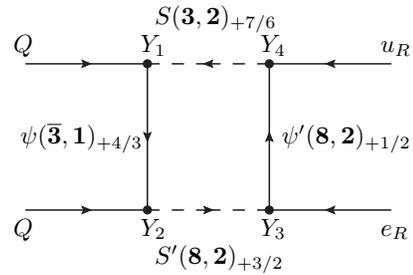}}
    \end{picture}
 \caption{An example, Model II,
 for one-loop decomposition of the proton
 decay operator $\mathcal{O}_{2}$.}
 \label{Fig:Model-1}
\end{figure}

Let us consider now a model without additional discrete symmetries,
which we categorized into the second class
in Section \ref{sect:intro}.
The full new particle content of the model is
\begin{gather}
\label{PartContII}
 \psi(\overline{\bf 3},{\bf 1})_{+4/3} = \psi_{+4/3} ,
 \quad
 S({\bf 3},{\bf 2})_{+7/6} = (S_{+2/3},S_{+5/3}),
 \nonumber
 \\
 \psi'({\bf 8},{\bf 2})_{+1/2}=(\psi^{\prime}_ 0,\psi^{\prime}_{+1}),
 \quad
 S'({\bf 8},{\bf 2})_{+3/2}=(S^{\prime}_{ +1},S^{\prime}_{ +2}),
\end{gather}
which corresponds to choose  the first row in Tab.~\ref{Tab:Summary-O2}
of the decomposition of the $ \mathcal{O}_{2}$ operator
with the parameter $\alpha=4/3$ for the electroweak $U(1)$ hypercharge.
For the colour representation, we take \#6 in Tab.~\ref{Tab:SU3-decom}.
The corresponding Feynman diagram of the 1-loop proton decay is shown
in Fig.~\ref{Fig:Model-1}. 
%
%

\begin{figure}
\unitlength=1cm
\begin{picture}(5.5,5)
\put(0,0){\includegraphics[width=5.3cm]{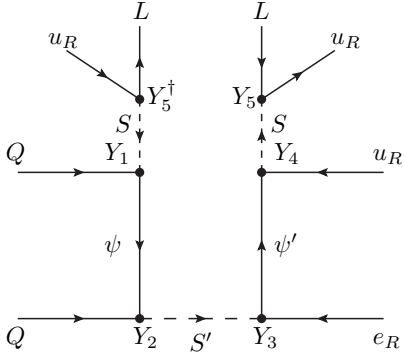}}
  \end{picture}
\caption{$d=12$ effective operator $Q Q u e \bar{u} L \bar{L} u$ at tree-level. }
\label{fig:diagd12}
\end{figure}

The symmetries allow to have an additional interaction 
\begin{align}
 \mathscr{L}_{2}
 =
 Y_{5}
 \overline{u_{R}}
 L
 {\rm i}\tau^{2}
 S
 +
 {\rm H.c.},
 \label{eq:L-decom}
\end{align}
which does not appear in the 1-loop proton decay diagram, shown in
Fig.~\ref{Fig:Model-1}. With this interaction, a $d=12$ effective
operator $Q Q u_{R} e_{R} \overline{u_{R}} L \overline{L} u_{R}$
appears at tree-level, as shown in Fig.~\ref{fig:diagd12}, which
causes 5-body proton decays such us $p \rightarrow e^+ e^- e^+ \pi^+
\pi^-$.
However, the decay modes induced from the $d=12$ operator
are sub-dominant;
Using Eq.~\eqref{eq:pdec}, for $Y_5 \sim 10^{-2}$, we can roughly
estimate the contribution of these modes to the proton total decay
width to be around 40 orders of magnitude smaller than the 2-body
proton decay induced by the $d=6$ effective operator $\mathcal{O}_{2}$
given through the 1-loop diagram in Fig.~\ref{Fig:Model-1}.

Combining the information listed in Tabs.~\ref{Tab:SU3-decom}
and \ref{Tab:Summary-O2}, one can find the coefficient of
the effective interaction $\mathcal{O}_{2}$ of proton decay as
\begin{align}
 \mathcal{C}_{2}
 =
 -\frac{8}{3}M_{\psi} M_{\psi'} I_{4}
 Y_{1}Y_{2}Y_{3}Y_{4},
 \label{eq:C1-Model-2}
\end{align}
and the effective interaction causes
only $p \rightarrow \pi^{0} e^{+}$.
%
As we have seen in Section \ref{sect:3.1},
the experimental bounds on the proton decay rate
require
Yukawa couplings of order $Y < \mathcal{O}(10^{-6})$
for the masses of the mediators at the TeV scale.

\begin{figure}[t]
\unitlength=1cm
\begin{picture}(7.8,5.5)
\put(0,0){\includegraphics[width=7.3cm]{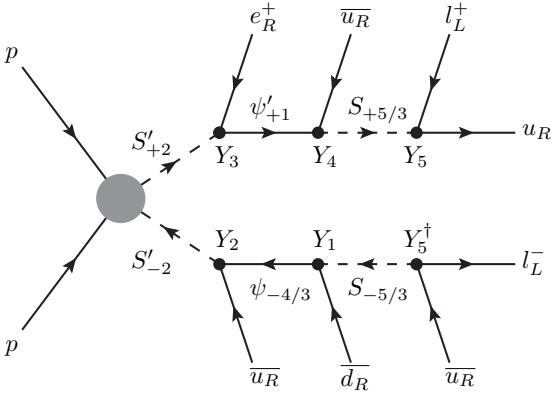}}
 \end{picture}
 \caption{Pair production and corresponding possible decays modes
 of the colour-octet scalar $S^{\prime}_{+2}$ at the LHC.}
\label{fig:LNV}
\end{figure}

Within this model it is possible to have signatures that violate the
lepton number by 1 unit, $\Delta L = 1$. This can be seen in the pair
production of the scalar $S^{\prime}_{ + 2} $ which has two possible
decay modes $S^{\prime}_{+ 2} \rightarrow l^+ l^+ 2j$ and
$S^{\prime}_{ + 2} \rightarrow l^+ 3j$, as shown in Fig.~\ref{fig:LNV}.
Therefore the pair production of the colour-octet
$S^{\prime}_{ +2}$ at the LHC might lead to the lepton number violating
(LNV) signal 3 lepton plus 5 jets $(l^+ l^- l^\pm 5j)$.
Observation of LNV through this process is only possible,
if $\Gamma(S^{\prime}_{ + 2} \rightarrow l^+ l^+ 2j )$ is
of similar order to $\Gamma(S^{\prime}_{ + 2} \rightarrow l^+ 3j)$,
since both final states are needed to establish
that LNV is indeed taking place. Similar order of these decay widths
are possible if $Y_1 Y_2 \sim Y_{3} Y_4$.  Observing this LNV process
also requires to have short enough decays so the decays of the
particles can be prompt.

This can be achieved for instance if we assume a mass hierarchy of the
particles, $M_{S^{\prime}} \gtrsim M_{\psi} (M_{\psi^\prime}) \gtrsim
M_{S} $ such that the pair production of $S^{\prime}_{ +2}$ leads to a
decay chain of 2-body decays as shown in Fig. \ref{fig:LNV}.
If this is the case, the two decay modes of $S^{\prime}_{ + 2}$ will be
$S^{\prime}_{ + 2} \rightarrow \psi^{\prime}_{ +1} l^+$ and
$S^{\prime}_{ + 2} \rightarrow \psi_{ + 4/3} j$, and its decay length
can be estimated as
\begin{align}
\label{decayL2}
 L_0(S^{\prime}_{ +2})
 \sim  10^{-2} \text{[m]}
 \frac{\left[10^{-6}\right]^2}{|Y_3|^2+|Y_4|^2 }
 \left[\frac{\text{TeV}}{M_{S^\prime}}\right].
\end{align}
Here, the choice for the Yukawa couplings $Y_3$ and $Y_4$ being order
$10^{-6}$ is motivated by the current proton decay experimental
bounds.

\begin{figure}[t] 
\unitlength=1cm
\hspace*{-1cm}
\begin{picture}(4.5,2)
\put(0,0){\includegraphics[width=4.5cm]{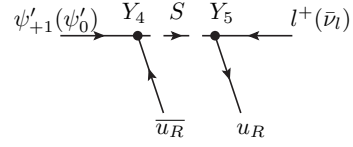}}
 \end{picture}
 \caption{ Decay channels of the colour-octet fermion $\psi^\prime$.
 }
 \label{fd-2}
\end{figure}

It is also possible to have long-lived particles at the LHC, which
are pair-produced.
Let us assume for instance that the colour-octet
fermion $\psi^\prime$ is slightly
lighter than the scalar $S$.
Then, the decay rates of the particles
$\psi^{\prime}_{ +1}\rightarrow l^+ j j$ and $\psi^{\prime}_{
  0}\rightarrow \nu_l j j$ can be estimated as (see Fig.~\ref{fd-2})
\begin{align}
 \Gamma(\psi^\prime
 \rightarrow
 l (\nu) jj)
 \sim
 \frac{|Y_{4} Y_{5}|^{2}}{512\pi^3}
 \left[
 \frac{M_{\psi^\prime}}{M_{S}}
 \right]^4
 M_{\psi^\prime},
\end{align}
which leads to the estimate of the decay length
\begin{align}
\label{decayL}
 L_0(\psi^\prime)
 \sim 30 \text{[m]}
 \left[\frac{10^{-6}}{|Y_{4}|}\right]^2
 \left[\frac{10^{-2}}{|Y_{5}|}\right]^2
 \left[\frac{\text{TeV}}{M_{\psi}}\right]
 \left[\frac{M_{S}}{M_{\psi^\prime}}\right]^4.
\end{align}
Here, the Yukawa coupling $Y_{5}$, which is not constrained by
the proton decay, has been set to be of the order $10^{-2}$.
Eq. (\ref{decayL}) shows that $\psi^{\prime }$, after being pair
produced at the LHC, will become a long lived particle. Since
$\psi^{\prime }$ has also colour, one can use
R-hadron searches at the LHC \cite{Alimena:2019zri, Aaboud:2019trc,
  Farrar:2010ps, Buccella:1985cs, CMS:2016ybj,Liu:2015bma}
to constrain it. 

In summary, in Model-II depending on the mass hierarchies of
the particles that appear in the proton decay diagram, one can
have either long-lived coloured/charged particles or prompt
decays. The latter would allow to establish experimentally
the existence of LNV.

\section{Conclusions}

We have studied systematically the 1-loop decomposition of the $d = 6$
$B + L$ violating operators involving only SM fields.  Our results are
listed in tabular forms, from which all possible 1-loop ultra-violet
completions of these operators involving fermions and scalars can be
constructed. We have briefly discussed how to use the information
provided to calculate all coefficients that enter the calculation of
the proton decay rate.

We then discussed that all models, in general, can be divided into two
distinct classes. Class-I models are those, for which the 1-loop
proton decay exists, but is not necessarily the dominant diagram for a
given model. Models in this class therefore need usually an additional
symmetry, such that tree-level contributions to proton decay become
forbidden. Class-II models are then simply those, for which the
particle content guarantees that the 1-loop diagram is automatically
the dominant contribution to proton decay.

We then turned to possible phenomenology of these models and discussed 
one example model from each class. Class-I models have an absolutely 
stable particle and thus proton decay can be connected to the dark 
matter in the universe. In the example we discussed, the same symmetry 
is responsible for 1-loop proton decay, dark matter and neutrino 
mass. The latter is also generated at 1-loop level, as in the 
scotogenic neutrino mass model. We have discussed existing constraints 
and possible LHC phenomenology of this model also briefly.

For the example model of class-II, we have discussed possible LHC
phenomenology. It has been shown that depending on the mass hierarchy
of the particles in the model, we can have particles that after being
pair produced at the LHC can decay promptly, leading to LNV signals,
or are long-lived coloured/charged particles. Signals without missing
energy, such as the LNV signals discussed, do not appear in Model-I
and so can be used to distinguish between these two classes of models.

\acknowledgments

M. H. was funded by Spanish grants FPA2017-90566-REDC (Red Consolider
MultiDark), FPA2017-85216-P and SEV-2014-0398 (MINECO/AEI/FEDER,
UE), as well as PROMETEO/2018/165 (Generalitat Valenciana). 
This work is supported in part by
European Union's Horizon 2020 research and innovation programme
under the 777419-ESSnuSB,
as well as by the COST Action CA15139 EuroNuNet.
T. O. acknowledges the support of
``Spanish Agencia Estatal de Investigaci\'{o}n'' (AEI)
through the grant ``IFT Centro de Excelencia Severo
Ochoa SEV-2016-0597'' and
the EU ``Fondo Europeo de Desarrollo Regional'' (FEDER) through the
project FPA2016-78645-P.
J. C. H. is supported by Chile grant Fondecyt No. 1161463.

\appendix
\section{Factors in Tables}

Here we write down 
the basis operators given in Eqs.~\eqref{eq:O1}-\eqref{eq:O5}
with all the indices of the SM gauge groups
and Lorentz spinors:
\begin{align}
 \mathcal{O}_{1}
 \equiv&
 \epsilon^{IJK}
 [
 (\overline{{d_{R}}^{c}})_{I\dot{a}}
 (u_{R})_{J}^{\dot{a}}
 ]
 [
 (\overline{Q^{c}})_{Ki}^{a}
 ({\rm i}\tau^{2})^{ij}
 (L)_{ja}
 ],
 \nonumber
 \\
 \mathcal{O}_{2}
 \equiv&
 \epsilon^{IJK}
 [
 (\overline{Q^{c}})_{Ii}^{a}
 ({\rm i}\tau^{2})^{ij}
 (Q)_{Jja}
 ]
 [
 (\overline{{u_{R}}^{c}})_{K \dot{a}}
 (e_{R})^{\dot{a}}
 ],
 \nonumber
 \\
 \mathcal{O}_{3}
 \equiv&
 \epsilon^{IJK}
 [
 (\overline{Q^{c}})_{Ii}^{a}
 ({\rm i}\tau^{2})^{ij}
 (Q)_{Jja}
 ]
 [
 (\overline{Q^{c}})_{Kk}^{b}
 ({\rm i}\tau^{2})^{kl}
 (L)_{l b}
 ],
 \nonumber
 \\
 \mathcal{O}_{4}
 \equiv&
 \epsilon^{IJK}
 [
 (\overline{Q^{c}})_{Ii}^{a}
 ({\rm i}\tau^{2} \tau^{d})^{ij}
 (Q)_{Jja}
 ]
 [
 (\overline{Q^{c}})_{Kk}^{b}
 ({\rm i}\tau^{2} \tau^{d})^{kl}
 (L)_{l b}
 ],
 \nonumber
 \\
 \mathcal{O}_{5}
 \equiv&
 \epsilon^{IJK}
 [
 (\overline{{d_{R}}^{c}})_{I \dot{a}}
 (u_{R})_{J}^{\dot{a}}
 ]
 [
 (\overline{{u_{R}}^{c}})_{K \dot{b}}
 (e_{R})^{\dot{b}}
 ],
\end{align}
where the different indices are introduced to 
describe the different representations:
$I,J,K \in \{1,2,3\}$ for a triplet under the colour $SU(3)$,
$i,j,k,l \in \{1,2\}$ for a doublet under the electroweak $SU(2)$, 
$d \in \{1,2,3\}$ on the Pauli matrices $\tau^{d}$
for a triplet under the $SU(2)$,
$a,b \in \{1,2\}$ for a left-handed 2-spinor
and
$\dot{a},\dot{b} \in \{\dot{1},\dot{2}\}$
for a right-handed 2-spinor.
The position of the indices also depends on the representation
of the field:
A lower $I$ for ${\bf 3}_{I}$
and an upper $I$ for $\overline{\bf 3}^{I}$.
The index on the $\overline{\bf 2}^{i}$ representation of $SU(2)$
can be lowered as ${\bf 2}_{i}$
with $({\rm i} \tau^{2})_{ij}$.
On the position of the spinor indices,
we follow the notation that is widely adopted in literature,
e.g. Ref.~\cite{Haber:1984rc};
the standard positions are determined as
$(\psi_{L})_{a}$ and $(\psi_{R})^{\dot{a}}$
and the contraction is taken as
$(\overline{\psi_{R}})^{a}(\psi_{L})_{a}$
and
$(\overline{\psi_{L}})_{\dot{a}} (\psi_{R})^{\dot{a}}$
to form Lorentz scalars.
%

The ordering of the field operators in the decomposed interactions
are determined as given in Eq.~\eqref{eq:L-decom-general}.
In order to make the Yukawa interactions
singlets under the colour $SU(3)$,
we plug
the total anti-symmetric tensors
($\epsilon^{IJK}$ and $\epsilon_{IJK}$),
the Gell-Mann matrices (${(\lambda^{A})_{I}}^{J}$),
and the Clebsch-Gordan (CG) coefficient matrices
($(T_{\bf 6})_{X}^{IJ}$ and
$(T_{\overline{\bf 6}})^{X}_{IJ}$)
into the interactions accordingly,
where
the index $A \in \{1 \cdots 8\}$ is
for an octet,
and
a lower (upper) $X \in\{1\cdots 6\}$ is
for ${\bf 6}$ ($\overline{\bf 6}$). 
For the contraction of the electroweak $SU(2)$ indices,
we use the anti-symmetric tensors
($({\rm i}\tau^{2})^{ij}$ and $({\rm i}\tau^{2})_{ij}$)
and
the Pauli matrices (${(\tau^{d})_{i}}^{j}$).
The CG matrices for the sextet representations are defined as
\begin{gather}
 (T_{\bf 6})_{1}^{IJ} =
 \begin{pmatrix}
  1 & &
  \\
  & 0 &
  \\
  & & 0
 \end{pmatrix}
 ,
 \quad 
 (T_{\bf 6})_{2}^{IJ} =
 \begin{pmatrix}
  0 & \frac{1}{\sqrt{2}} &
  \\
  \frac{1}{\sqrt{2}} & 0 &
  \\
  & & 0
 \end{pmatrix}
 \nonumber
 \\
 (T_{\bf 6})_{3}^{IJ} =
 \begin{pmatrix}
  0 & &
  \\
  & 1 &
  \\
  & & 0
 \end{pmatrix},
 \quad
  (T_{\bf 6})_{4}^{IJ} =
 \begin{pmatrix}
  0 & & \frac{1}{\sqrt{2}} 
  \\
  & 0 &
  \\
  \frac{1}{\sqrt{2}} & & 0
 \end{pmatrix}
 \nonumber
 \\
 (T_{\bf 6})_{5}^{IJ} =
 \begin{pmatrix}
  0 & & 
  \\
  & 0 & \frac{1}{\sqrt{2}} 
  \\
  & \frac{1}{\sqrt{2}} & 0
 \end{pmatrix},
 \quad
  (T_{\bf 6})_{6}^{IJ} =
 \begin{pmatrix}
  0 & &
  \\
  & 0 &
  \\
  & & 1
 \end{pmatrix},
\end{gather}
and $T_{\overline{\bf 6}}$ are defined in the same manner.

Let us demonstrate the operator projection
(=re-integrate out the mediator fields),
keeping all the indices,
i.e.,
we explicitly derive a basis operator(s) from a decomposition
with all the coefficients, signs and factors.
As an example, we take
the basis operator $\mathcal{O}_{1}$
and
decompose it with the mediators with 
\#4 in Tab.~\ref{Tab:SU3-decom}
\#12 for Tab.~\ref{Tab:Summary-O1},
i.e,
\begin{gather}
 \psi({\bf 3}, {\bf 3})_{\alpha},
 \quad
 S({\bf 8},{\bf 3})_{\alpha+1/3},
 \nonumber
 \\
 \psi'({\bf 6},{\bf 3})_{\alpha - 1/3}
 \quad
 S'({\bf 6}, {\bf 2})_{\alpha+1/6}.
\end{gather}
The Yukawa interactions are defined as
\begin{align}
  \mathscr{L}
 =&
 Y_{1}
 (\overline{{d_{R}}^{c}})_{I \dot{a}}
 {({\lambda^{\sf T}}^{A})^{I}}_{I'}
 ({\psi_{L}}^{c})^{I' d \dot{a}}
 S^{A d}
 \nonumber
 \\
 &+
 Y_{2}
 (\overline{{\psi_{L}}^{c}})_{J'}^{d' a}
 (T_{\bf 6})^{J'J}_{X}
 (Q)_{J i a}
 {({\tau^{\sf T}}^{d'})^{i}}_{i'}
 (S'^{\dagger})^{X i'}
 \nonumber
 \\
 &
 +
 Y_{3}
 (\overline{\psi'_{R}})^{X' f b}
 (L)_{j b}
 ({\rm i} \tau^{2} \tau^{f})^{jj'}
 S'_{X' j'}
 \nonumber
 \\
 &+
 Y_{4}
 \epsilon^{KLM}
 (\overline{{u_{R}}^{c}})_{K \dot{b}}
 {(\lambda^{A'})_{L}}^{N}
 (T_{\overline{\bf 6}})^{Y}_{NM}
 (\psi'_{R})_{Y}^{f' \dot{b}}
 (S^{\dagger})^{A' f'}
 \nonumber
 \\
 &+
 {\rm H.c.}
\end{align}
The effective proton decay operator 
resulting from the box diagram mediated by them
can be calculated as follows.
First, the mediator fields are contracted,
which give the propagators:
\begin{align}
 \mathscr{L}_{\text{eff}}
 =&
 Y_{1}Y_{2}Y_{3}Y_{4}
 {({\lambda^{\sf T}}^{A})^{I}}_{I'}
 (T_{\bf 6})^{J'J}_{X}
 \epsilon^{KLM}
 {(\lambda^{A'})_{L}}^{N}
 (T_{\overline{\bf 6}})^{Y}_{NM}
 \nonumber
 \\
 &\times
 {({\tau^{\sf T}}^{d'})^{i}}_{i'}
 ({\rm i} \tau^{2} \tau^{f})^{jj'}
 \Bigl\langle
 S^{A d}
 (S^{\dagger})^{A' f'}
 \Bigr\rangle
 \Bigl\langle
 S'_{X' j'}
 {(S'^{\dagger})}^{X i'}
 \Bigr\rangle
 \nonumber
 \\
 &\times
 (\overline{{d_{R}}^{c}})_{I \dot{a}}
 \Bigl\langle
 ({\psi_{L}}^{c})^{I' d \dot{a}}
 (\overline{{\psi_{L}}^{c}})_{J'}^{d' a}
 \Bigr\rangle
 (Q)_{J i a}
 \nonumber
 \\
 &\times
 (\overline{{u_{R}}^{c}})_{K \dot{b}}
 \Bigl\langle
 (\psi'_{R})_{Y}^{f' \dot{b}}
 (\overline{\psi'_{R}})^{X' f b}
 \Bigr\rangle
 (L)_{j b}
 \nonumber
 \\
 =&
 Y_{1}Y_{2}Y_{3}Y_{4}
 {({\lambda^{\sf T}}^{A})^{I}}_{I'}
 (T_{\bf 6})^{I'J}_{X}
 \epsilon^{KLM}
 {(\lambda^{A})_{L}}^{N}
 (T_{\overline{\bf 6}})^{X}_{NM}
 \nonumber
 \\
 &\times
 ({\rm i} \tau^{2} \tau^{d})^{ji'}
  {(\tau^{d})_{i'}}^{i}
 \int
 \frac{{\rm d}^{d} p}{(2\pi)^{d} {\rm i}}
 \frac{{\rm i}}{p^{2} - M_{S}^{2}}
 \frac{{\rm i}}{p^{2} - M_{S'}^{2}}
 \nonumber
 \\
 &\times
 (\overline{{d_{R}}^{c}})_{I \dot{a}}
 \frac{-{\rm i} p_{\rho} (\overline{\sigma}^{\rho})^{\dot{a}a}}
 {p^{2} - M_{\psi}^{2}}
 (Q)_{J i a}
 (\overline{{u_{R}}^{c}})_{K \dot{b}}
 \frac{{\rm i} p_{\sigma} (\overline{\sigma}^{\sigma})^{\dot{b}b}}
 {p^{2} - M_{\psi'}^{2}}
 (L)_{j b}.
\end{align}
Next, the $SU(3)$ and the $SU(2)$ indices are rearranged:
\begin{align}
 \mathscr{L}_{\text{eff}}
 =&
 Y_{1}Y_{2}Y_{3}Y_{4}
 \left[
 4 \epsilon^{IJK}
 \right]
 \left[
 - 3 ({\rm i} \tau^{2})^{ij}
 \right]
 \left[
 -\frac{1}{4} J_{4}
 \right]
 \nonumber
 \\
 &\times
 (\overline{{d_{R}}^{c}})_{I \dot{a}}
 (\overline{\sigma}^{\rho})^{\dot{a}a}
 (Q)_{J i a}
 (\overline{{u_{R}}^{c}})_{K \dot{b}}
 (\overline{\sigma}_{\rho})^{\dot{b}b}
 (L)_{j b}.
\end{align}
Here we arrived at the step shown in Eq.~\eqref{eq:SU3-projection}.
If necessary, the $SU(3)$ indices are renamed
so that they fit to the ordering in the corresponding basis operator.
This step may give an additional sign
(``$SU(3)$ sign'' in the tables):
\begin{align}
 \mathscr{L}_{\text{eff}}
 =&
 Y_{1}Y_{2}Y_{3}Y_{4}
 \left[
 4 \epsilon^{IJK}
 \right]
  \left[
 -1
 \right]
 \left[
 - 3 ({\rm i} \tau^{2})^{ij}
 \right]
 \left[
 -\frac{1}{4} J_{4}
 \right]
 \nonumber
 \\
 &\times
 (\overline{{d_{R}}^{c}})_{I \dot{a}}
 (\overline{\sigma}^{\rho})^{\dot{a}a}
 (Q)_{K i a}
 (\overline{{u_{R}}^{c}})_{J \dot{b}}
 (\overline{\sigma}_{\rho})^{\dot{b}b}
 (L)_{j b}.
\end{align}
Finally,
the Fierz transformation (rearrangement of the Lorentz indices)
is carried out:
\begin{align}
 \mathscr{L}_{\text{eff}}
 =&
 Y_{1}Y_{2}Y_{3}Y_{4}
 \left[
 4 
 \right]
 \left[
 -1
 \right]
 \left[
 - 3 
 \right]
 \left[
 -\frac{1}{4} J_{4}
 \right]
 \left[
 2
 \right]
 \nonumber
 \\
 &\times
 \epsilon^{IJK}
 (\overline{{d_{R}}^{c}})_{I \dot{a}}
 (u_{R})_{J}^{\dot{a}}
 (\overline{Q^{c}})_{K i}^{a}
 ({\rm i} \tau^{2})^{ij}
 (L)_{j a}
 \nonumber
 \\
 =&
 -6 Y_{1}Y_{2}Y_{3}Y_{4} J_{4} \mathcal{O}_{1}.
 \label{eq:Op-Projection-example}
 \end{align}

In this example, we obtain 4 for the $SU(3)$ coefficient,
$-$ for the $SU(3)$ sign,
$-3$ for the $SU(2)$ coefficient,
$2$ for the factor of the Fierz transformation for Lorentz indices,
and
$-J_{4}/4$ for the loop integral factor.
The loop integral factors $I_{4}$ and $J_{4}$
that appear in Tabs.~\ref{Tab:Summary-O1}-\ref{Tab:Summary-O5}
are defined as
\begin{align}
 I_{4} \equiv&
 \int \frac{{\rm d}^{4} k}{(2\pi)^{4}{\rm i}}
 \frac{1}{
 (k^{2} - M_{\psi}^{2})
 (k^{2} - M_{S}^{2})
 (k^{2} - M_{\psi'}^{2})
 (k^{2} - M_{S'}^{2})
 }
 \\
 J_{4} \equiv&
 \int \frac{{\rm d}^{4} k}{(2\pi)^{4}{\rm i}}
 \frac{k^{2}}{
 (k^{2} - M_{\psi}^{2})
 (k^{2} - M_{S}^{2})
 (k^{2} - M_{\psi'}^{2})
 (k^{2} - M_{S'}^{2})
 }
\end{align}
In the limit where all the mediator masses are identical,
the integrals converge to
\begin{align}
 I_{4} \rightarrow \frac{1}{16 \pi^{2}} \frac{1}{6} \frac{1}{M^{4}},
 \quad
 J_{4} \rightarrow -\frac{1}{16 \pi^{2}} \frac{1}{3} \frac{1}{M^{2}},
\end{align}
where $M$ is the common value of the masses.
All the information to reproduce the coefficient
of the effective operator $\mathcal{O}_{1}$
in Eq.~\eqref{eq:Op-Projection-example}
can be found in Tabs.~\ref{Tab:SU3-decom} and \ref{Tab:Summary-O1}.


\end{document}